\documentclass[aps,prd,amssymb,showpacs,floatfix,twocolumn,superscriptaddress,nofootinbib]{revtex4-1}
\usepackage{amsmath,amsfonts,graphics,epsfig,amssymb,bm}
\usepackage{multirow}
\usepackage{enumerate}
\usepackage{graphicx}
\usepackage [latin1]{inputenc}
\usepackage[colorlinks, citecolor=blue, anchorcolor=blue, linkcolor=blue]{hyperref}

\usepackage[doipre={doi:~}]{uri}

\begin{document}
%\captionsetup[figure]{labelformat={default},labelsep=period,name={FIG.}}

\title{The radiation emitted from axion dark matter in a homogeneous magnetic field, and possibilities for detection}
\author{Shuo Xu}
\affiliation{School of Physics, Sun Yat-sen University, Guangzhou, GuangDong, People's Republic of China}
\affiliation{Department of Astronomy, Tsinghua University, People's Republic of China}
\author{Siyu Chen}
\author{Hong-Hao Zhang}
\email[Corresponding author. ]{zhh98@mail.sysu.edu.cn}
\author{Guangbo Long}
\email[Corresponding author. ]{longgb@mail2.sysu.edu.cn}
\affiliation{School of Physics, Sun Yat-sen University, Guangzhou, GuangDong, People's Republic of China}

\begin{center}
\begin{abstract}
 We study the direct radiation excited by oscillating
axion (or axion-like particle) dark matter in a homogenous magnetic field and its detection scheme.
We concretely derive the analytical expression of the axion-induced radiated power for a cylindrical uniform magnetic field. In the long wave limit, the radiation power is proportional to the square of the B-field volume and the axion mass $m_a$, whereas it oscillate as approaching the short wave limit and the peak powers are proportional to the side area of the cylindrical magnetic field and $m_a^{-2}$. The maximum power locates at mass $m_a\sim\frac{3\pi}{4R}$ for fixed radius $R$. Based on this characteristic of the power, we discuss a scheme to detect the axions in the mass range $1-10^4$\,neV, where four detectors of different bandwidths surround the B-field. The expected sensitivity for $m_a\lesssim1\,\mu$eV under typical-parameter values can far exceed the existing constraints.
  %such improvements on the flux analysis lead to,  but have been neglected by relevant study %

\end{abstract}
\end{center}
\maketitle

\section{Introduction}
\label{sec:intro}

The axion originates from the Peccei-Quinn solution to the strong $CP$ problem. Its generalization, axionlike particles (ALPs, in the following referred to as axion), are predicted by many extensions of the Standard Model~\cite{Svrcek:2006yi}. These well-motivated pseudoscalar particles of electrically-neutral are also the prime candidates for the dark matter (DM)~\cite{Preskill1983,Abbott1983,Dine1983}.

 The interaction between the axion and the photon is described by the Lagrangian ${\cal L}_{a\gamma}=g_{a\gamma}{\bf E}\cdot{\bf B}\,a$, where ${\bf E}$ represents the electric field, ${\bf B}$ the magnetic field, $a$ the axion, and $g_{a\gamma}$ the coupling constant. If the axion is the DM particle, there is a oscillating axion field $a(t)$ in our Galactic halo. The external magnetic field ${\bf B}_{\rm e}$ with the axion field induces an extra source term in Maxwell's equations as the effective electric current density~\cite{Millar2017JCAP}
 \begin{equation}
{\bf J}_a(t)=g_{a\gamma}{\bf B}_{\rm e}\dot a(t).
\label{Ja}
\end{equation}
  Therefore, the energy of axion dark mater (ADM) can be converted to the electromagnetic field and the converted photons is expected to be observed as long as the B-field is strong enough. A number of experimental schemes have been designed to detect this axion-induced signal~\cite{Sikivie:2020zpn}.

At present, most of the existing or proposed experiments (haloscopes) take the advantage of the resonant enhancement on the EM modes drived by ${\bf J}_a(t)$~\cite{Sikivie:2020zpn}. For $1\,\mu {\rm eV} \lesssim m_{a}\lesssim 50\,\mu$eV, The cavity haloscopes (e.g.~ADMX~\cite{ADMX:2021nhd}, WISPDMX~\cite{WISPDMX}, CAPP~\cite{CAPP:2020utb}, HAYSTAC~\cite{Brubaker:2016ktl}, QUAX~\cite{Alesini:2019ajt}, GrAHal~\cite{Grenet:2021vbb}, RADES~\cite{CAST:2020rlf},ORGAN~\cite{Goryachev:2017wpw}) in strong magnetic fields are optimal strategy, where a narrow-band axion-induced excitation around each of the cavity resonances can be achieved~\cite{Sikivie:1983ip}. For smaller mass $m_{a}\lesssim 1\, \mu$eV, it can be explored by LC circuit~\cite{DMRadio,Cabrera2008,Sikivie:2013laa,Kahn:2016aff,Zhang:2021bpa,Chen:2021bgy} or radio-frequency cavity haloscopes~\cite{Berlin:2019ahk,Bogorad:2019pbu,Lasenby:2019prg, Berlin:2020vrk,Sikivie1009}. The former has been experimentally realized by SHAFT~\cite{Gramolin:2020ict}, ABRACADABRA~\cite{Ouellet:2018beu}, ADMX-SLIC~\cite{Crisosto:2019fcj} and the BASE Penning trap~\cite{Devlin:2021fpq}. For higher mass  $50\,\mu {\rm{eV}}\lesssim m_{a}\lesssim\,1\,$eV, the traditional experiment of resonance cavity is limited by the smaller volume required by the resonant enhancement and impractical scan rates for high mass. As a result, several novel detection techniques of resonance or constructive interference, such as topological insulators target~\cite{Marsh:2018dlj,Schutte-Engel:2021bqm}, plasma Haloscopes~\cite{Lawson:2019brd}, multiple-cavity~\cite{Aja:2022csb,Jeong:2020mtp,Melcon:2018dba,Baryakhtar:2018doz} and dielectric haloscopes~\cite{Caldwell:2016dcw,MADMAX:2019pub,Millar2017JCAP}, have been proposed to exploit the ADM of this mass band.

 Another type of strategy to probe the high-mass ADM is the dish-antenna~\cite{Horns:2012jf} or solenoidal haloscope~\cite{BREAD:2021tpx} without resorting to the resonant enhancement on the signal. The designed detector catch the axion-induced emission from the magnetized
 dish-antenna or cylindrical metal barrel surface and the detectable power is proportional to the area of the reflecting surface $A_{\rm ref}$~\cite{Horns:2012jf,Jaeckel:2013eha}.
These haloscopes compensate the loss of the resonance enhancement by increasing the volume of the B-field and can work in a broadband state. Note that, the dielectric haloscopes~\cite{Caldwell:2016dcw,MADMAX:2019pub,Millar2017JCAP} evolved from the dish-antenna scheme.  

In this work, we generalize the strategy of dish-antenna to a fully open case without a reflector in the B-field, which (or a similar case) was also mentioned in an preprint~\cite{Horns:2013ira}. We focus on investigating the direct radiation excited by the oscillating electric current density ${\bf J}_a(t)$, and derive the analytical expression of radiated power for a cylindrical uniform B-field concretely, which is more common in scientific detection instruments. Then, we discuss the possibility of detecting the ADM by means of the derived
 radiated power.

The plan of this paper is as follows.
In Sec.~\ref{secII}, we will derive the radiated power and the energy flux from axion conversion in a homogeneous magnetic field.
Then, in Sec.~\ref{secIII} we will discuss the experimental detection of the ADM. Finally, discussions and conclusions are presented in Sec.~\ref{sec:discussion} and~\ref{sec:conclusion} respectively.

\section{The radiated power from axion conversion}
\label{secII}
This section derives the radiated power and the energy flux from the axion-photon conversion in a homogeneous static B-field $\mathbf{B}_e$. We use the Heaviside-Lorentz units, in which $c=1$, $k_{\rm B}=1$, the permeability and permitivity of the vacuum $\mu_0= \epsilon_0 = 1$.
\subsection{Axion electrodynamics}

We assume the axion field $a(t) = {\rm{Re}} (a_0\, e^{\rm{i}\,(\mathbf{k}_a\cdot \mathbf{x} - \omega t)})$ $\simeq$ ${\rm{Re}} (a_0\,e^{-\rm{i}\,\omega t})$ since ADM particles move non-relativistically based on $k_a\ll\omega$ and $\omega=m_a$. The retarded potential can be obtained from solving Maxwell's equations with the only source term $\mathbf{J}_a$ in Eq.\,(\ref{Ja})~\cite{Sikivie:2020zpn}
\begin{equation}
\mathbf{A} (\mathbf{x}) ={1\over 4 \pi}
\int_V
{\mathbf{j}_a\,{\rm e}^{-{\rm i}\,kr} \over r}~dV^{\,\prime},
\label{A1}
\end{equation}
where $k = \omega$, $\mathbf{j}_a={\rm i}\,g_{a\gamma} \omega a_0 \mathbf{B}_{\rm e}$, and $\epsilon=\mu=1$. Here, $r=\vert\mathbf{x} - \mathbf{x}^{\,\prime}\vert>$\,0 is the distance from the observation location $\mathbf{x}$ to $\mathbf{x}^{\,\prime}$ of the magnetic field region. The range of integration is in the region extended by the B-field volume $V$.

The axion-induced radiated the energy-flux $\mathbf{S}$ for E-field $\mathbf{E}$ and B-field $\mathbf{B}$ is given by
 \begin{equation}
\langle \mathbf{S} \rangle\,=\frac{\rm Re(\mathbf{E}^{*}\times \mathbf{B})}{2}
,\,\,\, \mathbf{B}=\bigtriangledown\times \mathbf{A}, \,\,\, \mathbf{E}=\frac{\rm i}{\omega}\bigtriangledown\times \mathbf{B}.
\label{S}
\end{equation}
The bracket $\langle ... \rangle$ represents a time average.

For convenience, we calculate the radiated power from the axion conversion in the Fraunhofer radiated zone, i.e.,~$r>\frac{L^2\omega}{\pi}$ and $r\simeq\vert\mathbf{x}\vert\gg L$, where $L$ is the feature size of the B-field region.
  The retarded potential in Eq.\,(\ref{A1}) can be reduced to
\begin{equation}
\mathbf{A}(\mathbf{x}) = {{\rm e}^{{\rm i}\,kr} \over 4 \pi r}\,\mathbf{j}_a(\mathbf{k})
+ {\rm O}({1 \over r^2}),
\label{A2}
\end{equation}
where $\mathbf{j}_a(\mathbf{k})=\int_V
{{\rm e}^{-{\rm i}\,\mathbf{k}\cdot\,\mathbf{x}^{\,\prime}}\mathbf{j}_a}~dV^{\,\prime}$ is the Fourier transform of $\mathbf{j}_a$. The direction of the wave vector satisfies $\mathbf{n}=\mathbf{k}/k=\mathbf{r}/r$.

The averaged electromagnetic power $dP$ radiated for area d$\mathbf{s}$ in
direction $\mathbf{n}$ can be represented by $dP=\langle \mathbf{S}\rangle\cdot d\mathbf{s}$=$\langle \mathbf{S} \rangle\cdot\mathbf{n}\,r^2 d\Omega$. Then, the time-averaged power in the Fraunhofer radiated zone can be estimated as
\begin{eqnarray}
P &=&\int_\Omega \langle \mathbf{S} \rangle\cdot\mathbf{n}\,r^2\,d\Omega
\nonumber\\
&=&\int_\Omega{k \omega \over 32\pi^2}
\vert \mathbf{n} \times  \mathbf{j}_a (\mathbf{k})\vert^2\,d\Omega\nonumber\\
 &=&
\frac{g_{a\gamma}k \omega^3 \vert a_0\vert^2}{32\pi^2}\,\int_\Omega\,\bigg| \int_V\,\mathbf{n}\times\mathbf{B_{\rm e}}\,
{{\rm e}^{-{\rm i}\,\mathbf{k}\cdot\,\mathbf{x}^{\,\prime}}}dV^{\,\prime}
\bigg|^2\,d\Omega \nonumber\\
&=& {\rho_a g_{a\gamma}^2  B_{\rm e}^2 \omega^2 \over 16 \pi^2}
\int_{0}^{2\pi}d\varphi\int_{0}^{\pi} d\theta\, {\rm sin}^3\theta\,
\bigg| \int_V\,
{{\rm e}^{-{\rm i}\,\mathbf{k}\cdot\,\mathbf{x}^{\,\prime}}}dV^{\,\prime}
\bigg|^2 \nonumber\\& &  ~\
\label{kpow}
\end{eqnarray}
where $\rho_a=\frac{\omega^2 \vert a_0\vert^2}{2}$ is the energy density of the ADM, and we asumme $\vert\mathbf{n}\times\mathbf{B_{\rm e}}\vert=B_{\rm e}\,{\rm sin}\theta$ for the homogenous extra B-field (see,  Fig.\,\ref{B}).

Note that the excited E-field in the homogeneous B-field is also an oscillating field ${\bf E}_a(t)=-g_{a\gamma}{\bf B}_{\rm e}\,a(t)$~\cite{Millar2017JCAP}.

\subsection{The axion-induced radiated power for a cylindrical shape of the magnetic field}
\begin{figure}
%\begin{center}
\includegraphics[angle=360, width=80mm]{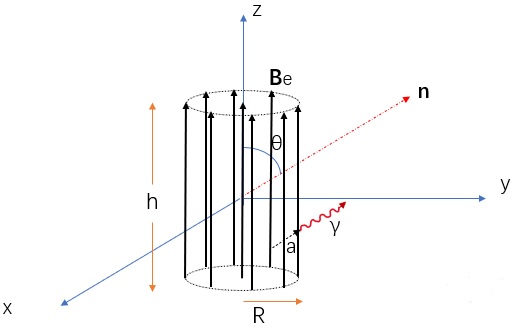}
\vspace{0.2in}
\caption{The schematic representation of a cylindrical magnetic field. The origin of x-y-z coordinate is located at the center of the cylindrical magnetic field which is parallel to the z axis. The vector $\mathbf{n}$ shows the direction of the wave vector of the radiated electromagnetic waves ($\gamma$) from axion ($a$) conversion, and its angle to the $z$ axis is denoted by $\theta$. The height and the radius of the cylindrical B-field is denoted by $h$ and $R$, respectively.}
%\end{center}
\label{B}
\end{figure}

We study the axion-induced radiation in a cylindrical uniform B-field
 with height $h$ and radius $R$, as shown in Fig.\,\ref{B}.
 The time-averaged radiated power can be further simplified by integrating over the volume of the B-field in Eq.\,(\ref{kpow}). We calculate $\int_V\,
{{\rm e}^{-i\,\mathbf{k}\cdot\,\mathbf{x}^{\,\prime}}}dV^{\,\prime}$ in cylindrical coordinate for $\mathbf{x}^{\,\prime}$ and in spherical coordinate for $\mathbf{k}$ as
\begin{eqnarray}
 I_V &=&
\int_{-\frac{h}{2}}^{\frac{h}{2}}{\rm exp}(-{\rm i}k{\rm cos}\theta z) dz\,\int_{0}^{2\pi}d\phi\int_{0}^{R}d\rho\,{\rm exp}(-{\rm i} \rho k {\rm sin}\theta \nonumber\\
& &({\rm cos}\varphi {\rm cos}\phi+{\rm sin}\varphi {\rm sin}\phi))\nonumber\\
&=& \frac{2{\rm sin}(\frac{h}{2}k{\rm cos}\theta)}{k{\rm cos}\theta}\int_{0}^{R} d\rho\int_{0}^{2\pi}d\phi\,{\rm e}^{-{\rm i} \rho k {\rm sin}\theta {\rm cos}(\varphi-\phi)} \nonumber\\
&=& \frac{2{\rm sin}(\frac{h}{2}k{\rm cos}\theta)}{k{\rm cos}\theta}\,\int_{0}^{R}d\rho\int_{0}^{2\pi}d\phi\,{\rm e}^{-{\rm i} \rho k {\rm sin}\theta {\rm cos}(\phi)},
\label{pow}
\end{eqnarray}
where we have used the column symmetry of the radiation and have chosen $\varphi$=0. By means of the
 zero-order Bessel-function ${\rm J}_0(x)=\frac{1}{2\pi}\int_{0}^{2\pi}{\rm exp}(-{\rm i} x {\rm cos}(\phi))d\phi$, the representation of $I_V$ can be reduced to
\begin{eqnarray}
 I_V &=& \frac{2{\rm sin}(\frac{h}{2}k{\rm cos}\theta)}{k{\rm cos}\theta}\,\int_{0}^{R}2\pi{\rm J}_0(\rho k {\rm sin}\theta)\rho\, d\rho \nonumber\\
 &=& \frac{2{\rm sin}(\frac{h}{2}k{\rm cos}\theta)}{k{\rm cos}\theta}\frac{2\pi}{(k {\rm sin}\theta)^2}\int_{0}^{Rk{\rm sin}\theta}(\rho k{\rm sin}\theta){\rm J}_0(\rho k {\rm sin}\theta)\nonumber\\
 & &d(k\rho {\rm sin}\theta)\nonumber\\
 &=& \frac{2{\rm sin}(\frac{h}{2}k{\rm cos}\theta)}{k{\rm cos}\theta}\,2\pi R^2\frac{{\rm J}_1(Rk{\rm sin}\theta)}{Rk{\rm sin}\theta},
\label{pow1}
\end{eqnarray}
where we have used the relation of $\int x{\rm J}_0(x)=x{\rm J}_1(x)$ in the last step of the derivation. The integral over $\rho$ and $\phi$ above implies the Fraunhofer diffraction of a circular hole.
Therefore, one can obtain the time-averaged radiated power for the ``cylindrical'' B-field as
\begin{eqnarray}
 P &=& {\rho_a g_{a\gamma}^2  B_{\rm e}^2 \omega^2 \over 16 \pi^2}
\int_{0}^{2\pi}d\varphi\int_{0}^{\pi}d\theta\,{\rm sin}^3\theta\,
|I_V|^2\nonumber\\
 &=&{2\pi \rho_a g_{a\gamma}^2  B_{\rm e}^2 R^4}\int_{0}^{\pi}{\rm sin}^3\,\theta \Big(\frac{{\rm sin}(\frac{h}{2}\omega{\rm cos}\theta)}{{\rm cos}\theta}\Big)^2\,\nonumber\\
 &&\Big(\frac{{\rm J}_1(R\omega{\rm sin}\theta)}{R\omega{\rm sin}\theta}\Big)^2 d\theta.
\label{pow2}
\end{eqnarray}

\begin{figure}[htbp]
  \centering
    \includegraphics[width=0.35\textwidth]{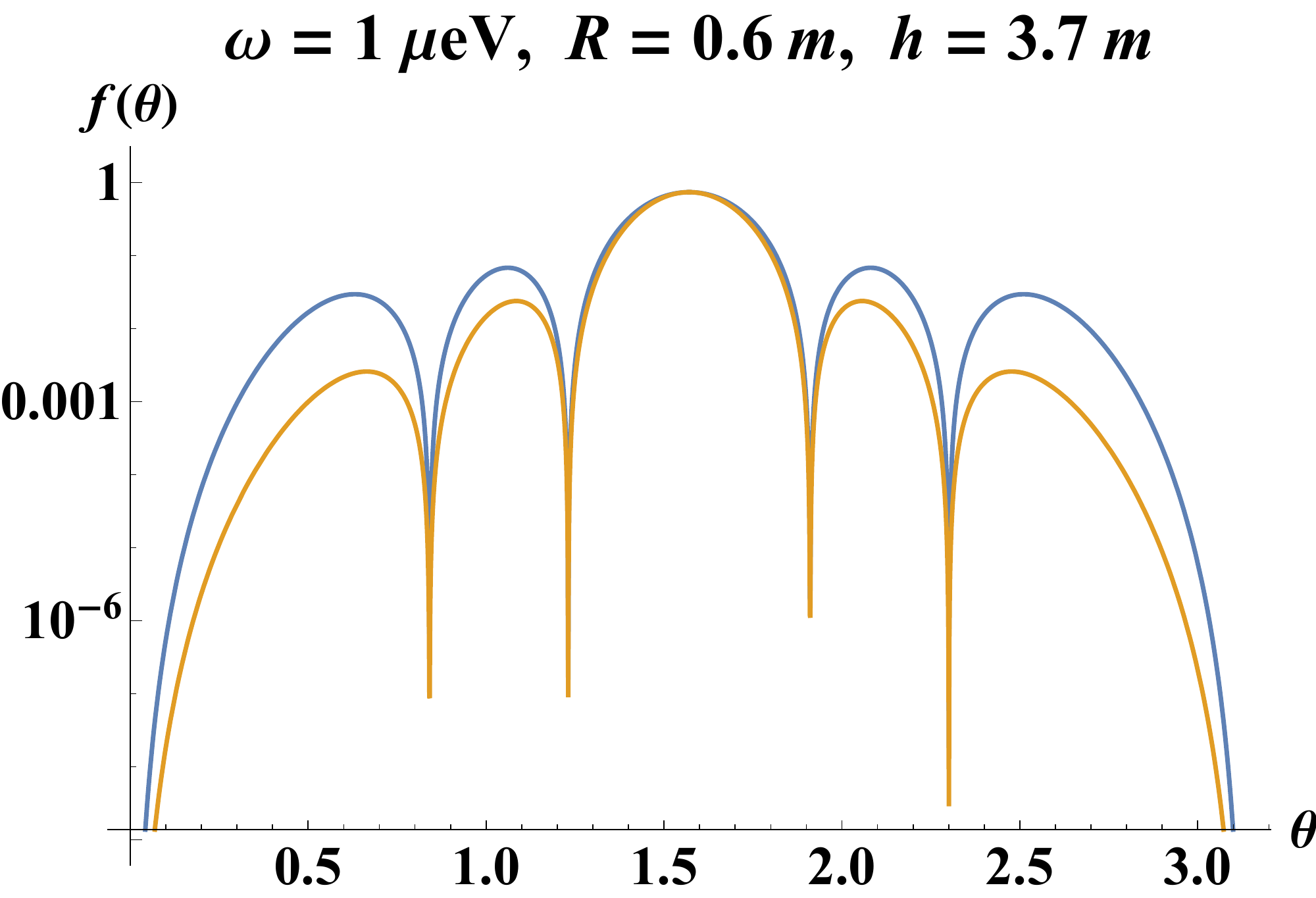}\\
     \includegraphics[width=0.35\textwidth]{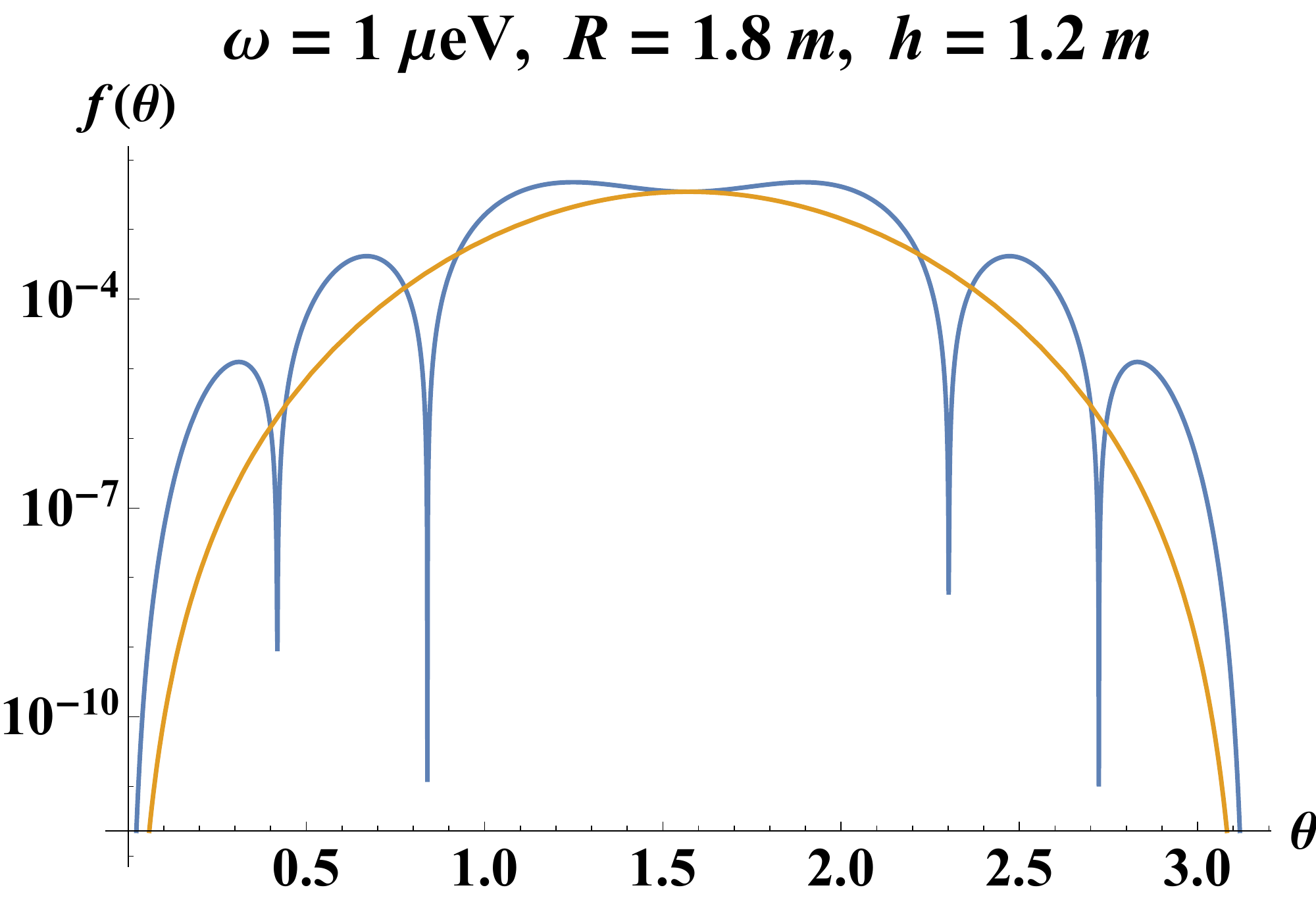}
    \includegraphics[width=0.35\textwidth]{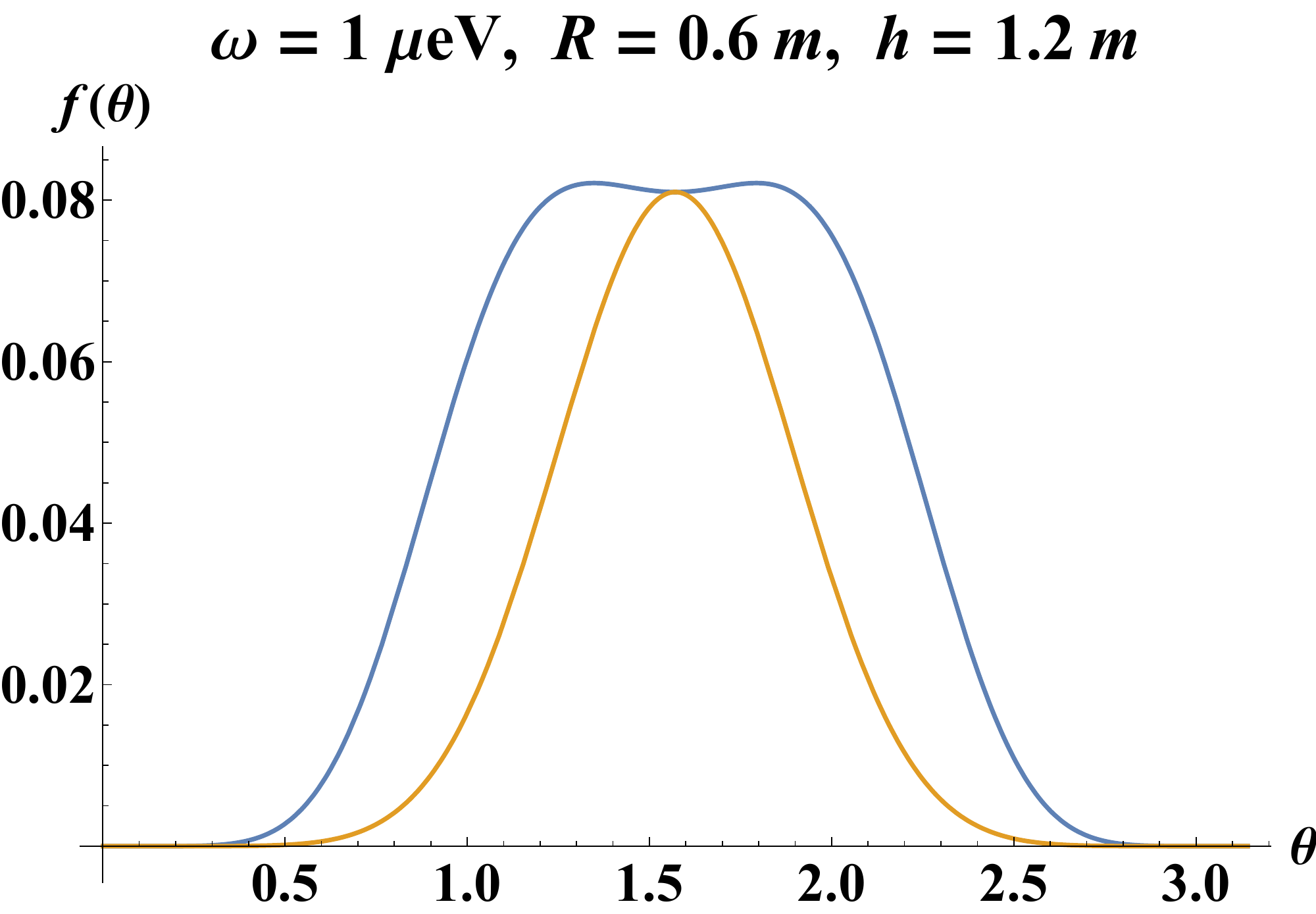}
    \caption{The integral functions ($f(\theta)$) in Eq.\,(\ref{pow2}) for different sizes of the B-field volume, where the yellow lines represent the approximated results with taking $\theta=\pi/2$ in the aperture diffraction function $\big(\frac{{\rm J}_1(R\omega{\rm sin}\theta)}{R\omega{\rm sin}\theta}\big)^2$. Top panel: $h=3\frac{2\pi}{\omega}\approx3.7\,$m, $R=\frac{\pi}{\omega}\approx0.6\,$m, and the ordinate is shown in logarithmic form; Middle panel: $h=\frac{2\pi}{\omega}, R=3\frac{\pi}{\omega}$, and the ordinate is shown in logarithmic form; Bottom panel: $h=\frac{2\pi}{\omega}=2R$. }
    \label{fig:f}
\end{figure}
Now, the integral above cannot be further calculated analytically and we need to study the integral function to approximate it.  Fig.\,\ref{fig:f} shows the numerical plots of the integral function in Eq.\,(\ref{pow2}) for different sizes of the B-field volume, where the yellow lines represent a approximation with taking $\theta=\pi/2$ in the ``aperture diffraction function'' $\big(\frac{{\rm J}_1(R\omega{\rm sin}\theta)}{R\omega{\rm sin}\theta}\big)^2$. The ordinates of the top and middle panels are shown in logarithmic form. We can find that the radiation is concentrated around the horizontal direction ($\theta=\frac{\pi}{2}$), which is similar to the principal maximum of Fraunhofer diffraction. The secondary peak number in the radiation angular distribution is $N=2(N_h+ N_R-2)$ when $N_h=\frac{h \omega}{2\pi}$ and $N_R=\frac{R\omega}{\pi}$ are integers. When $h>\frac{2\pi}{\omega}$ and $R>\frac{\pi}{\omega}$, the radiation is more concentrated horizontally. The approximations are more accurate for $h\gtrsim2R$, which means the dominance of the longitudinal radiation over the overall radiation, as shown in the top panel.

Inspired from the above analysis, we can use the delta function to approximate the longitudinal integral function as

\begin{eqnarray}
 P&=&{2\pi \rho_a g_{a\gamma}^2  B_{\rm e}^2 R^4}\int_{0}^{1}-{\rm sin}^3\theta \pi\frac{{\rm sin^2}(\frac{\omega h}{2}{\rm cos}\theta)}{\pi \frac{\omega h}{2}{\rm cos}^2\theta} \frac{\omega h}{2}\nonumber\\
 &&\Big(\frac{{\rm J}_1(R\omega{\rm sin}\theta)}{R\omega{\rm sin}\theta}\Big)^2 \frac{d({\rm cos}\theta)}{{\rm sin}\,\theta} \nonumber\\
 &\simeq& {2\pi^2 \rho_a g_{a\gamma}^2  B_{\rm e}^2 R^4}\int_{0}^{1}-{\rm sin}^3\,\theta \delta({\rm cos}\theta) \frac{\omega h}{2}\nonumber\\
 &&\Big(\frac{{\rm J}_1(R\omega{\rm sin}\theta)}{R\omega{\rm sin}\theta}\Big)^2 \frac{d({\rm cos}\theta)}{{\rm sin}\,\theta}\nonumber\\
 &=& {\pi^2 \rho_a g_{a\gamma}^2  B_{\rm e}^2 R^4 h \omega}
 \Big(\frac{{\rm J}_1(R\omega)}{R\omega}\Big)^2
\label{pd}.
\end{eqnarray}
To ensure the validity of the $\delta$ approximation, $\frac{\omega h}{2}$ is required to be sufficiently large compared to the other variate or parameter of the internal function, such as $\rm cos\theta$ and $\omega R$. Here, we require $\frac{\omega h}{2}>1$ (the maximum of ${\rm cos}\theta$) to ensure the horizontal focusing of the radiation and $h\gtrsim2R$ to ensure the dominance of the longitudinal radiation, as analyzed above.

We further expand the Bessel function for $\omega R>1$ so that Eq.\,(\ref{pd}) is reduced to
\begin{eqnarray}
 P&\simeq& 2\pi \rho_a g_{a\gamma}^2  B_{\rm e}^2 \frac{R h}{\omega^2}
 {\rm cos}^2(R\omega-\frac{3\pi}{4}),\,\,(\frac{\omega h}{2}\gtrsim\omega R>1) . \nonumber\\
\label{pJ}
\end{eqnarray}
We are more interested in the extremum (resonance point) of the power $P$, around which the $\delta$ approximation is also more valid. The power at the extreme points is given by
\begin{eqnarray}
P &=&\rho_a g_{a\gamma}^2  B_{\rm e}^2 \frac{A}{\omega^2},\,\big(\frac{\omega h}{2}\gtrsim R\omega=(n+\frac{3}{4}) \pi, n=0,1,2\cdot\cdot\cdot\big)\nonumber \\
&=&6.9\times10^{-23}{\rm W}\,\frac{\rho_a}{0.3\, \rm GeV\,cm^{-3}}\frac{h}{2\,\rm m}\frac{R}{\rm m}\Big(\frac{B_{\rm e}}{10\,\rm T}\Big)^2 \nonumber \\
 &&\Big(\frac{\rm \mu eV}{m_a}\Big)^2\Big(\frac{g_{a\gamma}}{10^{-14}\rm GeV^{-1}}\Big)^2,\,\,\,\big(h\gtrsim 2R,  \nonumber \\
 &&5.1 \frac{R}{\rm m}\frac{m_a}{\mu \rm eV}=(n+\frac{3}{4})\pi, \, n=0,1,2\cdot\cdot\cdot\big)\,\,\,,
\label{Pr}
\end{eqnarray}
where $A=2\pi Rh$ is the side area of the cylinder. The extreme radiated power is proportional to the surface area of B-field region and is inversely proportional to the square of the frequency, which is the same as the disk antenna experiment~\cite{Horns:2012jf}.

\begin{figure}[htbp]
  \centering
    \includegraphics[width=0.35\textwidth]{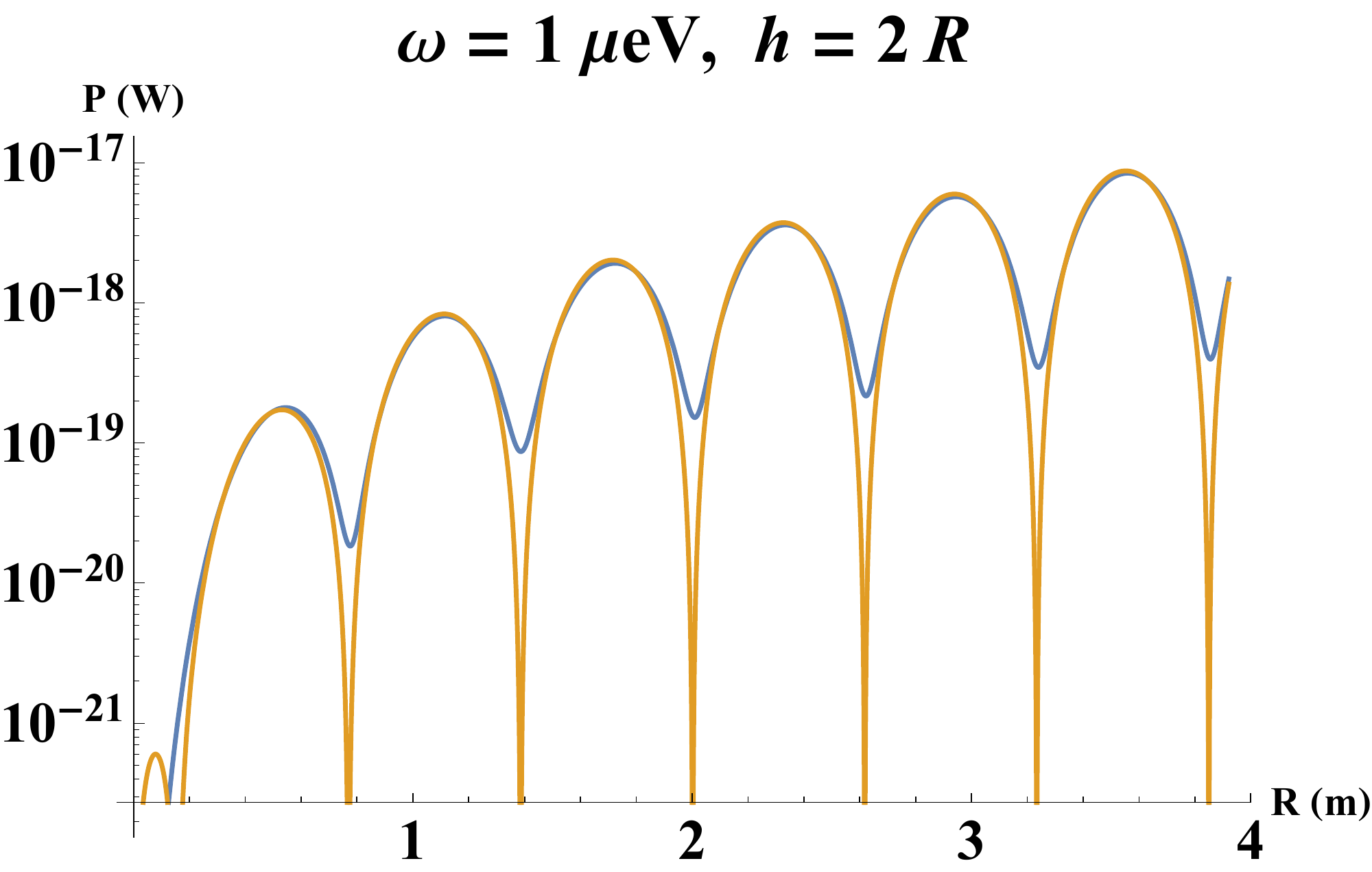}\\
    \includegraphics[width=0.35\textwidth]{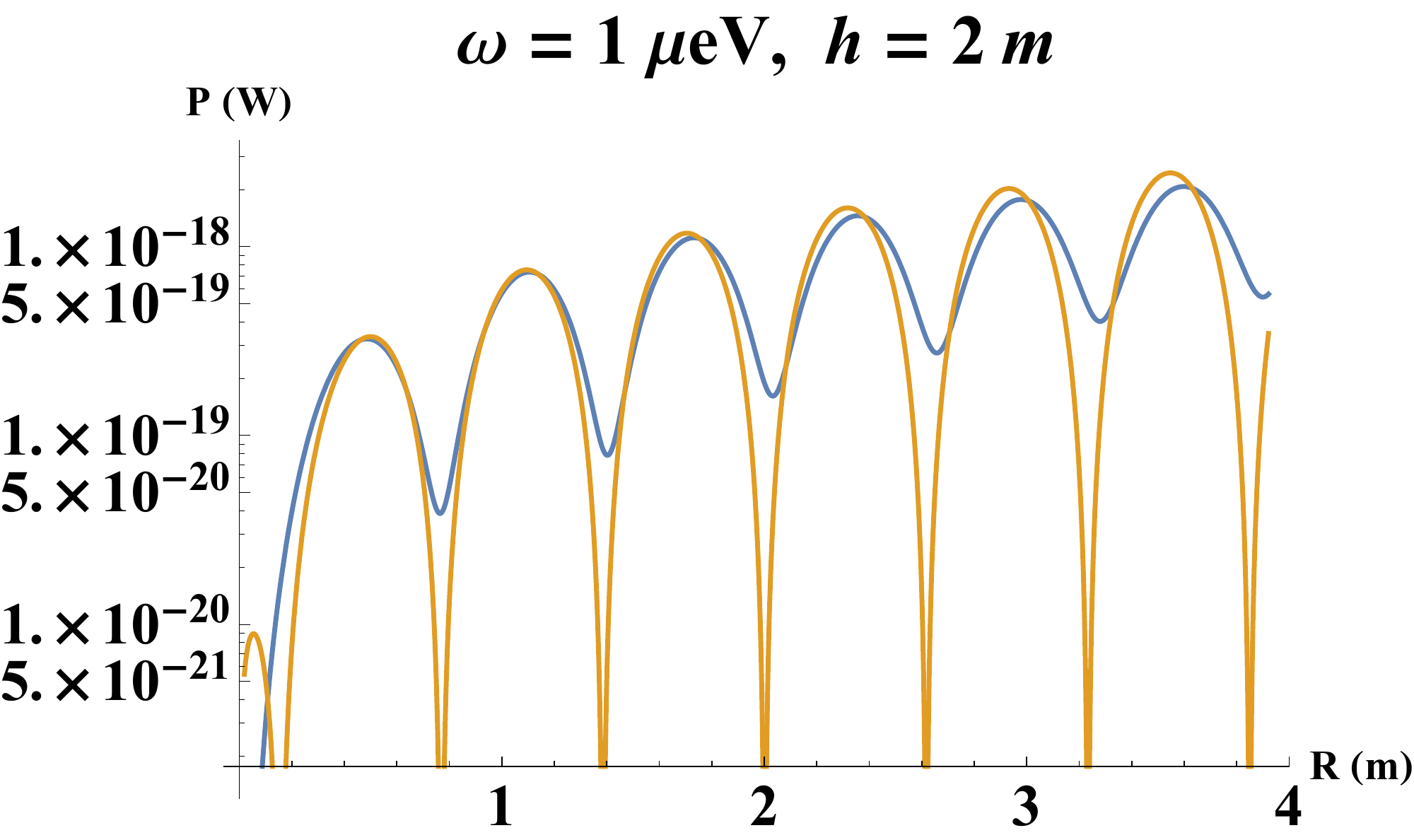}
     \includegraphics[width=0.35\textwidth]{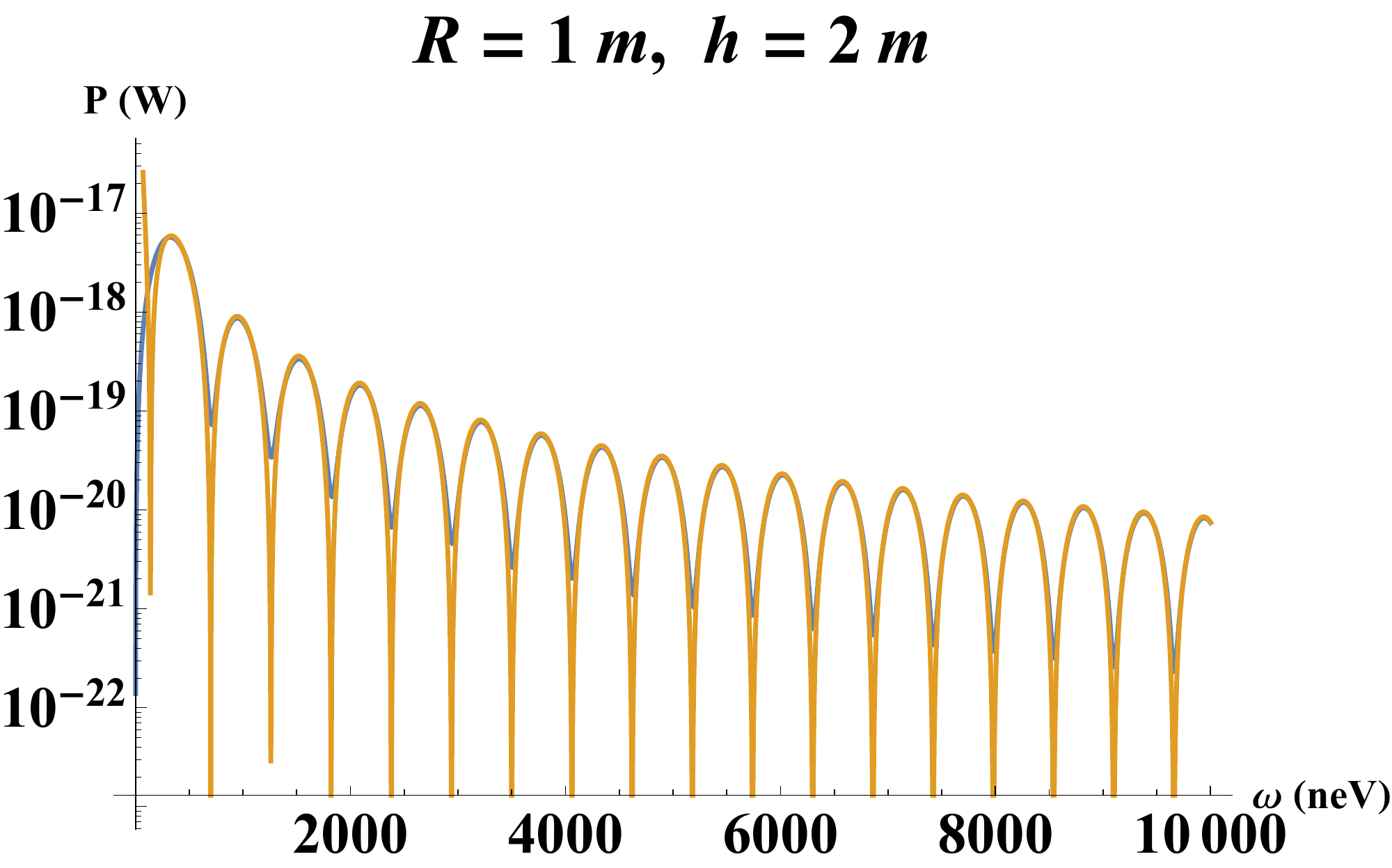}
    \caption{The comparison between the radiated power (blue line) resulted from numerical integral according to Eq.\,(\ref{pow2}) and its approximation (yellow line) according to Eq.\,(\ref{pJ}). Top panel: P as a function of the radius $R$ for $h=2 R$ and $\omega=1 \mu\,$eV; Middle panel: Like the case of the top panel, but $h$ is fixed to 2 m; Bottom panel: P as a function of the frequency for $h=2\,$m and $R=1\,$m. The approximation is good around the peaks when $\frac{\omega h}{2}\gtrsim\omega R>1$. When $\omega R<1$, the approximation fails as shown in all figures and also for the case $h<2R$ in the middle panel. The other parameter values are chosen as $g_{a\gamma}=10^{-12}\,{\rm GeV^{-1}}$, $B=10\,{\rm T}$, and $\rho_{a}=0.3\,{\rm GeV cm^{-3}}$.}
    \label{fig:ph}
\end{figure}

\begin{figure}[htbp]
  \centering
    \includegraphics[width=0.35\textwidth]{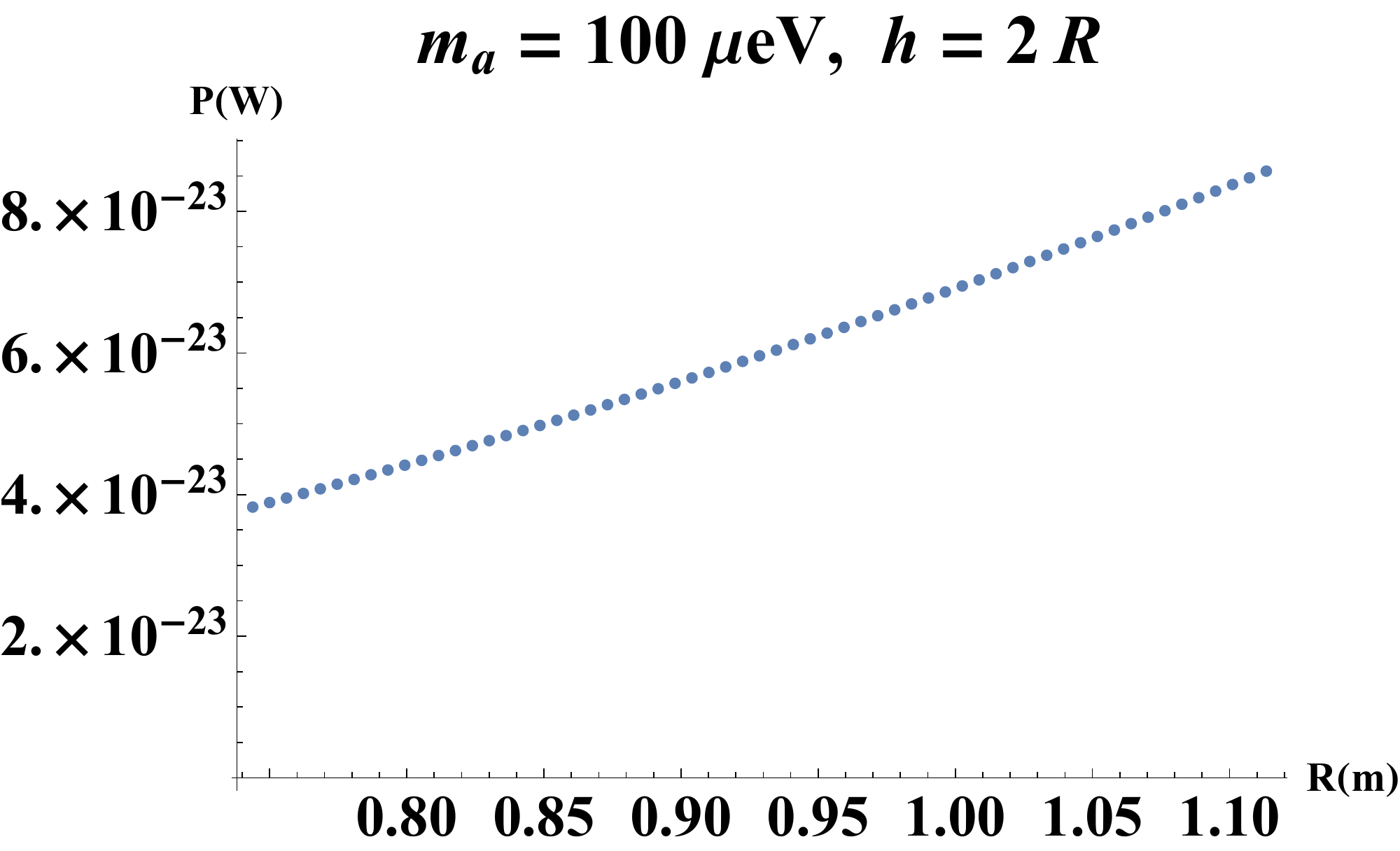}
    \caption{The radiated power at extreme points
  varies with R according to Eq.\,(\ref{Pr}). The other parameter values are taken as in Fig.\,\ref{fig:ph}.}
    \label{fig:pr}
\end{figure}

Fig.\,\ref{fig:ph} shows the comparison between the radiated power (blue line) resulting from the numerical integral according to Eq.\,(\ref{pow2}) and its approximation (yellow line) according to Eq.\,(\ref{pJ}). The approximation is good around the peaks when $\frac{\omega h}{2}\gtrsim\omega R>1$. However, when $\omega R<1$, the approximation fails at small $R$ or $\omega$ as shown in all three figures, and the middle panel shows a similar case of $h<2 R$ when $R>2\,$m.

For a fixed B-field volume, the maximum of the radiated power is the first peak at $\omega=\frac{3}{4R} \pi$ ($R\omega=(n+\frac{3}{4}) \pi>1$, n=0), before which the power $P(\omega)$ is an increasing function of the frequency $\omega^2$, and after which decreasing as $\frac{1}{\omega^2}$. This feature helps us find the best measurement frequency for a fixed B-field volume.

Fig.\,\ref{fig:pr} shows the radiated power of Eq.\,(\ref{Pr}) at extreme points which varies with R. The spacing between the extreme points is very small, which means that detecting such peak powers for $m_a=100\,\mu$eV requires the centimeter-level accuracy of the magnetic field device.

\begin{figure}[htbp]
  \centering
    \includegraphics[width=0.35\textwidth]{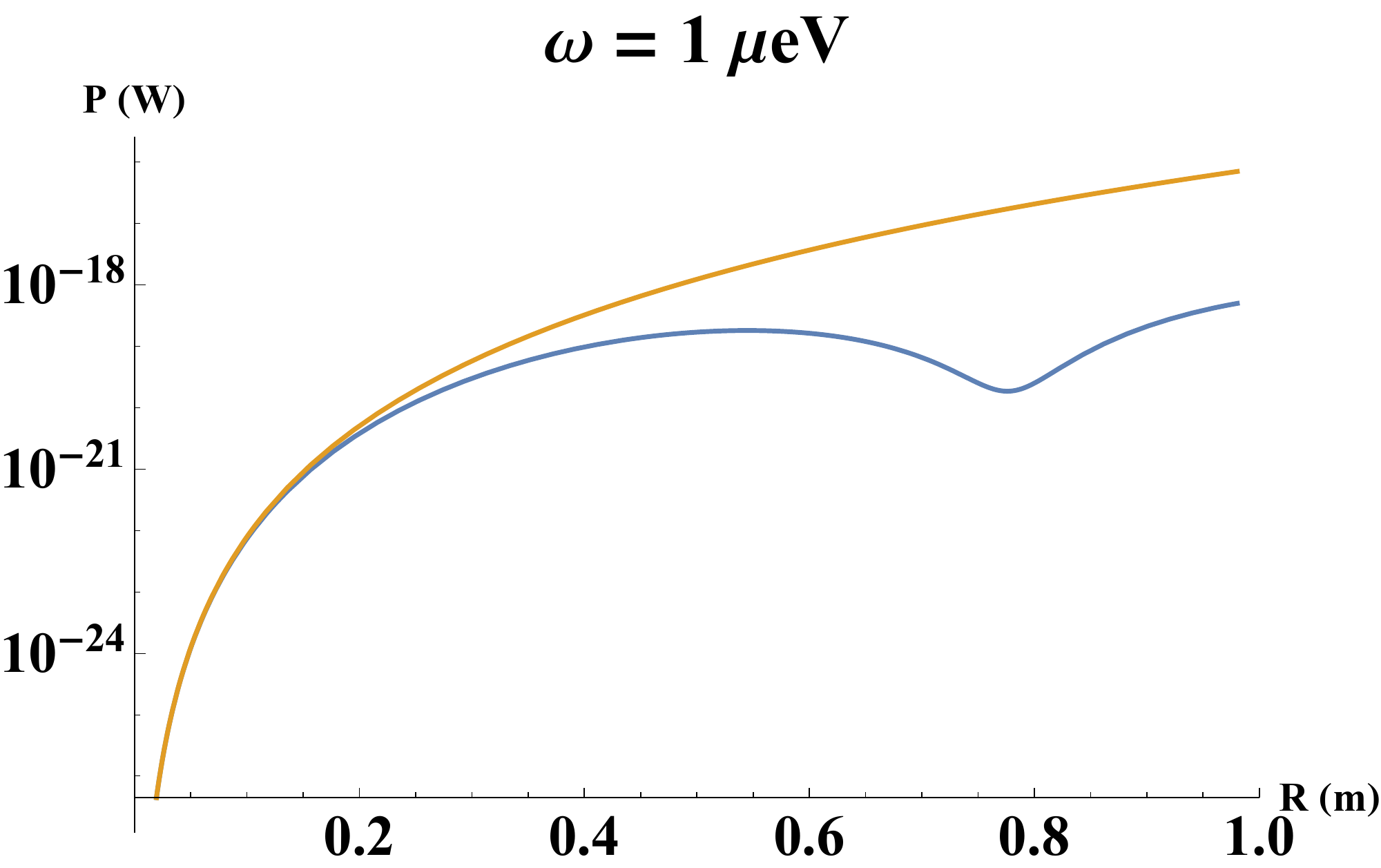}\\
     \includegraphics[width=0.35\textwidth]{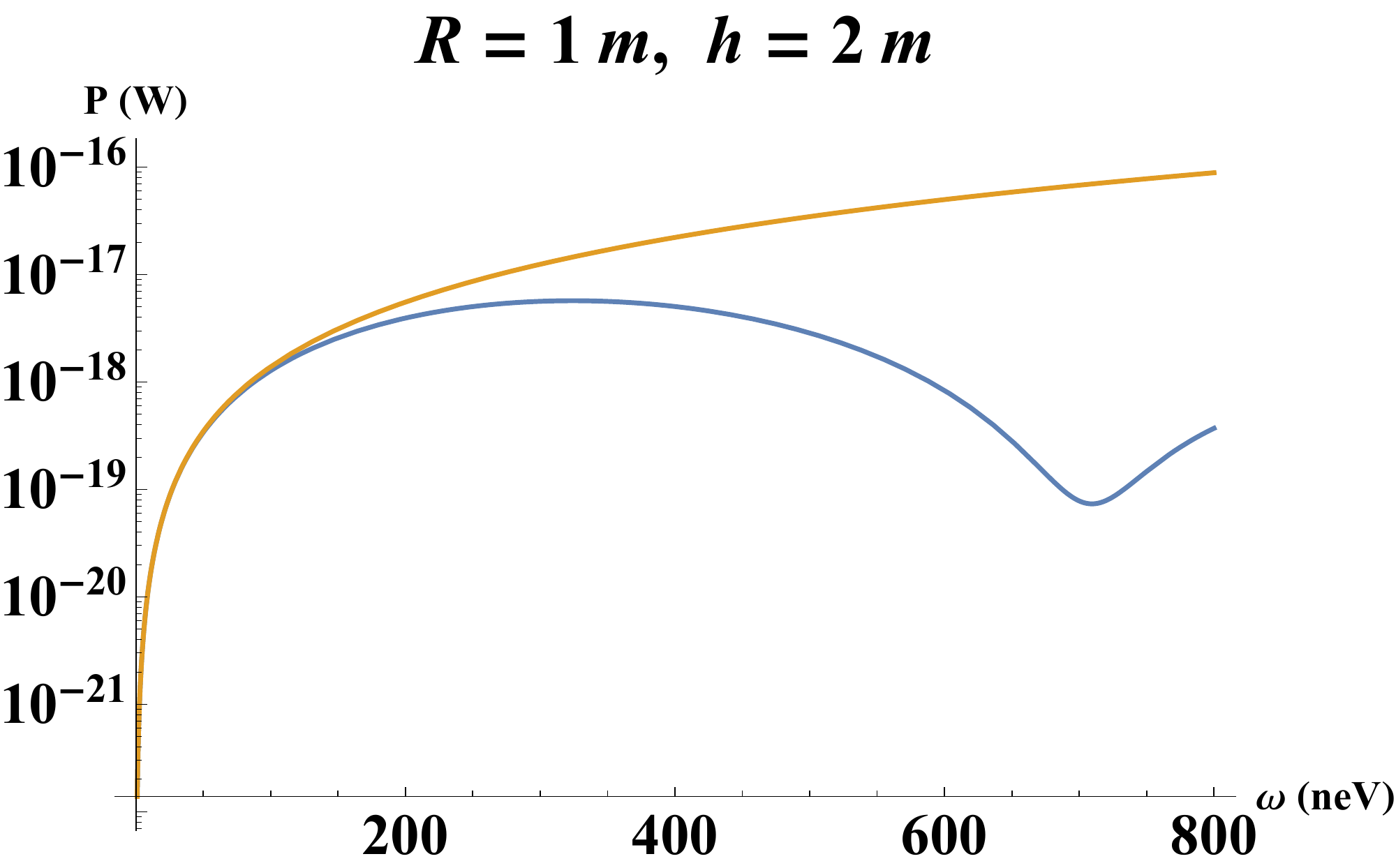}
    \caption{The comparison between the radiated power (blue line) according to Eq.\,(\ref{pow2}) and its approximation in the long wave limit (yellow line) as Eq.\,(\ref{Pl}). Top panel: $P$ as a function of the radius for $\omega=1 \mu\,$eV; Bottom panel: $P$ as a function of the frequency for $h=2\,$m and $R=1\,$m. The approximation is consistent with the numerical results when the condition $\omega R\ll1$ is satisfied. The other parameter values are chosen as $g_{a\gamma}=10^{-12}\,{\rm GeV^{-1}}$, $B=10\,{\rm T}$, and $\rho_{a}=0.3\,{\rm GeV cm^{-3}}$.}
    \label{fig:pl}
\end{figure}

The results are obtained in the short wave approximation for the radiated power. The result of the long wave approximation is easy to obtain under the approximation of ${{\rm e}^{-{\rm i}\,\mathbf{k}\cdot\,\mathbf{x}^{\,\prime}}}\simeq1$, that is

\begin{eqnarray}
P &=& {\rho_a g_{a\gamma}^2  B_{\rm e}^2 \omega^2 \over 16 \pi^2}
\int_{0}^{2\pi}d\varphi\,\int_{0}^{\pi}d\theta\, {\rm sin}^3\theta\,
\bigg| \int_V\,dV^{\,\prime}
\bigg|^2 ,\nonumber\\
&= &{\rho_a g_{a\gamma}^2  B_{\rm e}^2 V^2 \omega^2 \over 6 \pi},~\ (R\omega\ll1,\,h\omega\ll1)\nonumber \\
&=&7.8\times10^{-23}{\rm W}\,\frac{\rho_a}{0.3 \rm GeV\,cm^{-3}} \Big(\frac{B_{\rm e}}{10\rm T}\Big)^2 \Big(\frac{m_a}{\rm 100\,neV}\Big)^2\nonumber \\
 &&\Big(\frac{g_{a\gamma}}{10^{-14}\rm GeV^{-1}}\Big)^2\Big(\frac{h}{\rm 2m}\Big)^2\big(\frac{R}{\rm m}\big)^4,\,\,\,(0.51 \frac{R}{\rm m}\frac{m_a}{100 \rm neV} \nonumber \\
 &&\ll1,\,0.51 \frac{h}{\rm m}\frac{m_a}{100 \rm neV}\ll1).
\label{Pl}
\end{eqnarray}
The radiated power is proportional to the square of the B-field volume and the frequency, and is independent of the shape of the homogenous B-field region. Fig.\,\ref{fig:pl} (yellow line) shows the power in Eq.\,(\ref{Pl}) as a function of $R$ and $\omega$, respectively. Blue lines represent the radiated power (blue line) according to Eq.\,(\ref{pow2}). We can find the approximation is consistent with the numerical results when the condition $\omega R\ll1$ is satisfied.

This calculation pattern for the radiated power under the homogenous ``cylindrical magnetic field'' above can be generalized to the case with a B-field of cubic volume.
\subsection{The energy flux}
 The radiated power is independent of the distance to the radiation source, but the energy flux depends. In this part, we give the formula of the energy flux , which may be used to detect the axion-induced radiation.

 The electric and magnetic fields can be derived with  Eq.\,(\ref{A1}) and Eq.\,(\ref{S}) as
 \begin{eqnarray}
  \mathbf{B}&=&-{\rm i}\frac{a_0 \omega g_{a\gamma}B_e}{4\pi}\left(I_{1y}\mathbf{e}_x-I_{1x}
  \mathbf{e}_y\right) \label{equ:B}\\
  \mathbf{E}&=&\frac{a_0 g_{a\gamma}B_e}{4\pi}\big(I_{2xz}\mathbf{e}_x+I_{2yz}-(I_{2x^2}+I_{2y^2}
  +2I_{0})\mathbf{e}_y\big)\nonumber\\
  &&
   \label{equ:E}
\end{eqnarray}
 with x-y-z coordinate basis vector $\mathbf{e}_x$, $\mathbf{e}_y$, and $\mathbf{e}_z$. The integrals $I$ are defined as
\begin{eqnarray}
  I_0&=&\int_{V'}\frac{{\rm i}\omega r -1}{r^3}e^{{\rm i}\omega r}dV' \\
  I_{1x}&=&\int_{V'}\frac{{\rm i}\omega r -1}{r^3}e^{{\rm i}\omega r}(x-x')dV' \\
  I_{2x^2}&=&\int_{V'}\frac{3-3i\omega r-\omega^2r^2}{r^5}e^{{\rm i}\omega r}(x-x')^2dV' \\
  I_{2xz}&=&\int_{V'}\frac{3-3i\omega r-\omega^2r^2}{r^5}e^{{\rm i}\omega r}(x-x')(z-z')dV', \nonumber\\
         &&
\end{eqnarray}
where
\[
r=\sqrt{(x'-x)^2+(y'-y)^2+(z'-z)^2}
\]
 is the distance from the observation position ($\mathbf{x}$) to the source point ($\mathbf{x}'$). Then, we can obtain the average energy flux according to Eq.\,(\ref{A1})
 \begin{eqnarray}
  S_x&=&\frac{\rho_ag_{a\gamma}^2B_e^2}{16\pi^2\omega}\textrm{Re}\left[{\rm i}I_{1x}
  (I^\ast_{2x^2}+I^\ast_{2y^2}+2I^\ast_0)\right] \label{equ:numSx} \\
  S_y&=&\frac{\rho_ag_{a\gamma}^2B_e^2}{16\pi^2\omega}\textrm{Re}
  \left[{\rm i}I_{1y}
  (I^\ast_{2x^2}+I^\ast_{2y^2}+2I^\ast_0)\right]  \label{equ:numSy}\\
  S_z&=&\frac{\rho_ag_{a\gamma}^2B_e^2}{16\pi^2\omega}\textrm{Re}
  \left[{\rm i}
  (I^\ast_{2xz}I_{1x}+I^\ast_{2yz}I_{1y})\right].  \label{equ:numSz}
\end{eqnarray}
The component of the flux is defined as $S_{x,y,z}=\langle \mathbf{S} \rangle\cdot \mathbf{e}_{x,y,z}$\,.
They can be rewritten in the international unit, e.g.,\,for $S_x$, that is
\begin{gather}
  S_x=1.76\times10^{-20}\textrm{W}/\textrm{m}^2\frac{\rho_a}
  {0.3\textrm{GeV}/\textrm{cm}^3}\left(\frac{g_{a\gamma}}{10^{-12}
  \textrm{GeV}^{-1}}\right)^2\nonumber\\
\frac{100\textrm{neV}}{m_a}\left(\frac{B_e}{10\textrm{T}}\right)^2
\textrm{Re}
\left[{\rm i}\frac{I_{1x}}{\rm m}(I^\ast_{2x^2}+I^\ast_{2y^2}+2I^\ast_0)\right]. \label{equ:numSx0}
\end{gather}
\begin{figure}[htbp]
  \centering
    \includegraphics[width=0.35\textwidth]{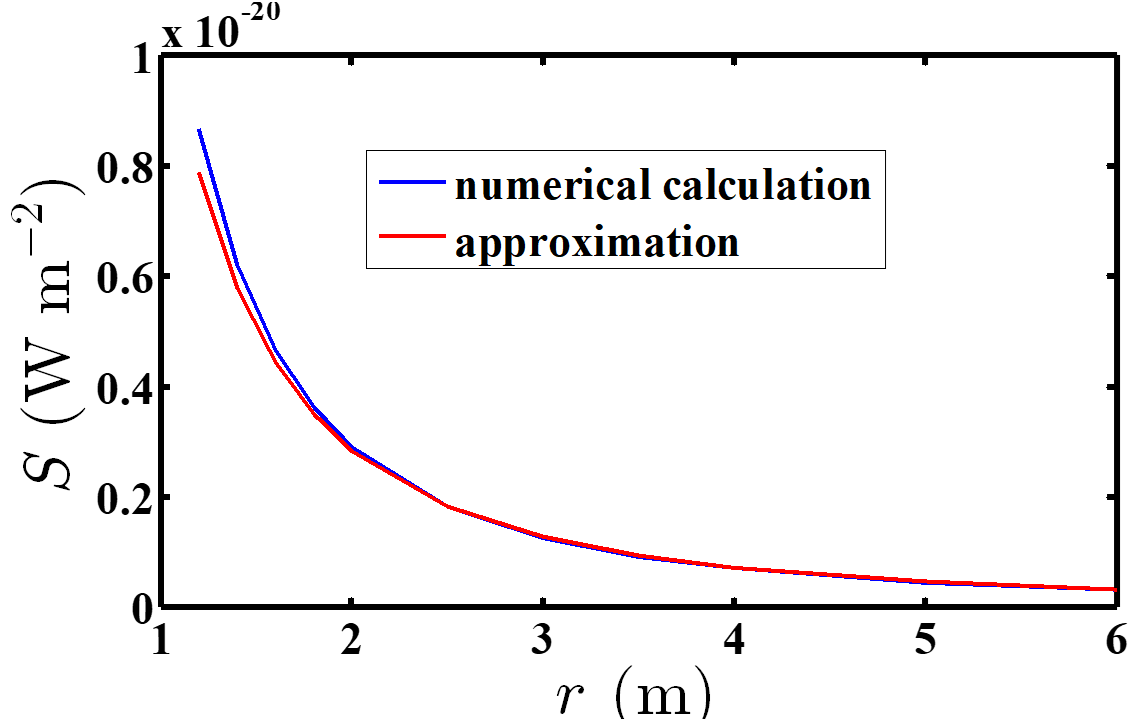}
    \caption{The comparison between the energy flux (blue line) according to Eq.\,(\ref{equ:S0}) and its approximation in the long wave and far-field limit (red line) as Eq.\,(\ref{equ:numSx}-\ref{equ:numSz}). The energy flux $S$ as a function of the observation distance for $\omega=100\,$neV, $g_{a\gamma}=10^{-12}{\rm GeV^{-1}}$, $B=10\,{\rm T}$, and $\rho_{a}=0.3\,{\rm GeV cm^{-3}}$. The approximation is consistent with the numerical result when the condition $|\mathbf{x}|\gg| \mathbf{x}'|$ is satisfied.}
    \label{fig:S}
\end{figure}

When the observation is located at the far-field region ($|\mathbf{x}|\gg| \mathbf{x}'|$) and the wave length is much larger than the size of the radiation source ($\lambda\gg| \mathbf{x}'|$), we can easily get the magnitude of the flux from Eq.\,(\ref{kpow}) as
\begin{eqnarray}
 S&=& {\rho_a g_{a\gamma}^2  B_{\rm e}^2 \omega^2 V^2\over 16 \pi^2r^2}{\rm sin}^2\theta\nonumber\\
 &=&2.3\times10^{-25}\textrm{W}/\textrm{m}^2\frac{\rho_a}
  {0.3\textrm{GeV}/\textrm{cm}^3}\left(\frac{g_{a\gamma}}{10^{-12}
  \textrm{GeV}^{-1}}\right)^2\nonumber\\
&&\left(\frac{m_a}{100\textrm{neV}}\right)^2\left(\frac{B_e}{10\textrm{T}}\right)^2
\left(\frac{V}{\rm m^3}\right)^2 \left(\frac{100\rm m}{r}\right)^2 {\rm sin}^2\theta\,.
\label{equ:S0}
\end{eqnarray}

Fig.\,\ref{fig:S} shows the comparison between the energy flux (blue line) according to Eq.\,(\ref{equ:numSx}-\ref{equ:numSz}) and its approximation in the long wave and the far-field limit (red line) as Eq.\,(\ref{equ:S0}). For $\omega=100\,$neV, when $r\gtrsim3\,$m, the condition $|\mathbf{x}|\gg| \mathbf{x}'|$ is satisfied, and the approximation is consistent with the numerical result. For higher frequency or axion mass, the integral function in Eq.\,(\ref{equ:numSx}-\ref{equ:numSz}) oscillates much faster, some special numerical techniques are required and it will be left to subsequent research.

\section{The experimental detection}
\label{secIII}
\begin{figure}[htbp]
  \centering
    \includegraphics[width=0.4\textwidth]{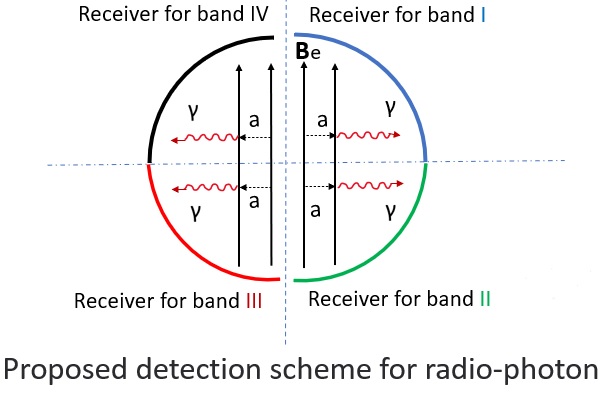}
    \caption{The profile map of a detection scheme for radio-microwave radiated photons ($\gamma$). The blue dashed lines represent the central axis. Four detectors worked on different bandwidths surround the magnetic field to detect the axion($a$)-induced radiation with four bandwiths: Band\,I (1-11\,neV), Band\,II (11-111\,neV), Band\,III (0.111-1.11$\mu$eV), Band IV (1.11-11.11\,$\mu$eV). The receiver can be dish antennas. Each detector with a specific bandwidth can consist of two identical self-detectors, as a small volume is conducive to lowering the operating temperature.}
    \label{fig:scheme}
\end{figure}

 We discuss the possible to detect the ADM-induced radiation for the homogenous ``cylindrical B-field''.

The radiated power increases with the volume or the side area of the magnetic field, which however must be finite. For a fixed B-field volume or the side area of the cylinder, the maximum of the power $P(\omega)$  occurs at the first peak determined by $\omega\simeq\frac{3}{4R} \pi$ ($R\omega=(n+\frac{3}{4}) \pi>1$, n=0), before which the power $P(\omega)$ is an increasing function of the frequency $\omega^2$, and after which decreasing as $\frac{1}{\omega^2}$, e.g., see the bottom panel of Fig.\,\ref{fig:ph}. If the maximum allowable radius $R$ is 2.5\,m, the maximum value of the $P(\omega)$ occurs at $\gtrsim$200\,neV.

Therefore, we attempt to focus on the bandwidth of $1-10^4$\,neV, which corresponds to the frequencies $0.24\,{\rm MHz}-24\,{\rm GHz}$. At this bandwidth, the typical detection level with techniques of the radio-astronomical measurement is $\lesssim10^{-22}$\,W~\cite{Horns:2013ira}. For typical parameters as in Eq.\,(\ref{Pr}) and (\ref{Pl}), the detectable axion-photon coupling $g_{a\gamma}$ can reach about $10^{-14}\,{\rm GeV^{-1}}$ which is allowed by the present observation except for a very narrow mass band excluded by ADMX~\cite{ADMX:2021nhd}. It seems to be promising to detect the radiation from such axions on experiment.

  The thermal noise of the detecting instrument is the main noise.
  Within a bandwidth of $\Delta\nu$, the ratio of the signal to one $\sigma$ fluctuation in the noise, i.e,\, the signal to noise ratio (SNR), is given by Dicke's radiometer equation~\cite{Dicke:1946glx}:
\begin{equation}
{\rm SNR} = {P_{\rm signal} \over T_{\rm n}} \sqrt{{\Delta t\over \Delta\nu}}
\label{radio}
\end{equation}
where $T_{\rm n}$ represents the total noise temperature and $P_{\rm signal}$ is the ADM-induced radiated power.

\subsection{broadband searches}
\begin{figure*}[t]
  \centering
    \includegraphics[width=0.65\textwidth]{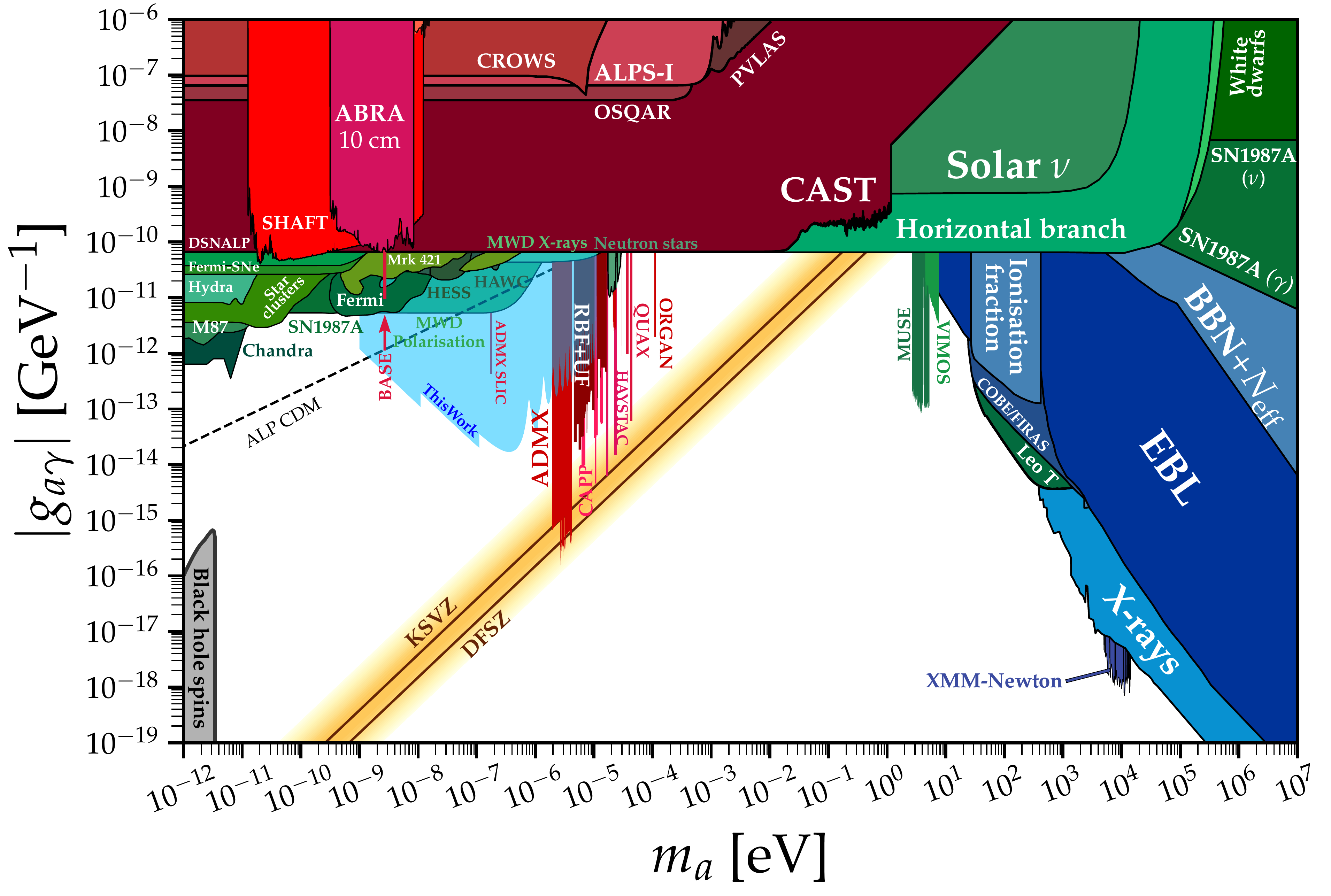}
    \caption{The expected sensitivity (in shades of light blue) for our detection schemes shown in Fig.\,\ref{fig:scheme}.
    we assumes the signal-to-noise ratio SNR=5. The height and radius of the cylindrical magnetic field are 1\,m and 2\,m, respectively. The detection time $\Delta t=3\,$years for each detector, total noise temperature $T_{\rm n}=1\,$K, $B=10\,{\rm T}$, and ADM energy density $\rho_{a}=0.3\,{\rm GeV cm^{-3}}$ are also assumed. The dark shading shows existing constraints. The lines of KSVZ~\cite{Kim:1979if,Shifman:1979if} and DFSZ~\cite{Dine1981,Zhitnitsky:1980tq} together orange strip are benchmark axion predictions. Limit data for other experiments and plotting scripts are available at Ref.\,\cite{Hare}. The dashed black line marks the upper bound on ALP DM~\cite{Arias:2012az}.}
\label{fig:sensitivity}
\end{figure*}
As we can see in Fig.\,\ref{fig:ph} and \ref{fig:pl}, the radiation peaks present wide, especially at low frequencies, since our scheme gives up the resonance enhancement on the ADM-induced signal.
 The measurement bandwidth $\Delta \nu$ of our scheme mainly depends on detector technology. We can use dish antennas as the detectors, which are widely used in radio astronomy. For modern detectors of radio astronomy, the bandwidths $\Delta \nu$ can be
in excess of 1\,GHz and the spectral resolutions can be better than $10^6$~\cite{Horns:2013ira}. We conservatively divide the interested detection bandwidth $1-10^4\,$neV into four section: Band\,I (1-11\,neV), Band\,II (11-111\,neV), Band\,III (0.111-1.11$\mu$eV), Band\,IV (1.11-11.11\,$\mu$eV). Since the radiometry device of cylinder is very large and has symmetry with respect to the x-y plane and z-axial, we can use four detectors with different bandwidths surrounding the magnetic field to detect the radiation at the Band\,I-IV simultaneously . Each detector occupies about a quarter of a sphere in space, see Fig.\,\ref{fig:scheme}.
The ideal detectable power for each detector is a quarter of total power
 \begin{equation}
P_{\rm signal} = \frac{1}{4} P.
\label{Eq:psignal}
\end{equation}

Note that the detectors should closely surround the magnetic field to avoid radiation energy leaking due to the diffraction effect. Each detector with a specific bandwidth can consist of two identical self-detectors, as a small volume is conducive to lowering the operating temperature.
\subsection{The expected sensitivity}
Using Eq.\,(\ref{pow2}), (\ref{radio}) and (\ref{Eq:psignal}),
the expected sensitivity for our detection scheme (Fig.\,\ref{fig:scheme}) is shown in shades of light blue in Fig.\,\ref{fig:sensitivity}. The detection time is assumed to $\Delta t=3\,$years, SNR=5, total noise temperature $T_{\rm n}=1\,$K, $B=10\,{\rm T}$, and $\rho_{a}=0.3\,{\rm GeV cm^{-3}}$. The height and radius of the cylindrical magnetic field are 1\,m and 2\,m, respectively.

In general, the sensitivity curve shows a shape with low middle and high ends, which is mainly determined by the radiated power. The strongest limitation can reaches $g_{a\gamma}\simeq10^{-14}\rm GeV^{-1}$ at about 500\,neV, as the maximum of the radiation power occurs at this mass point. The dark shading shows existing constraints. Our sensitivity is far beyond existing constraints except for ADMX which operates in the narrow bandwidth mode. If our scheme also only focus on a very narrow bandwidth in which the frequency corresponding to the maximum power is located by adjusting the radius $R$ of the B-field to satisfy $\omega\sim\frac{3}{4R} \pi$, the sensitivity is expected to increase by an order of magnitude.

\section{Discussion}
\label{sec:discussion}
We mainly discuss the comparison of our scheme with other experiments, especially the dish-antenna experiment that is similar to ours and the high mass axion detections.

\begin{figure}[htbp]
  \centering
     \includegraphics[width=0.4\textwidth]{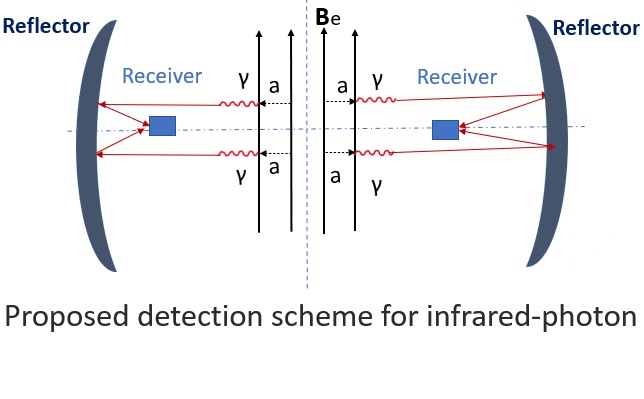}
    \caption{The profile map of detection scheme for infrared (about meV, corresponds to 0.24\,THz) radiated photons. The blue dashed lines represent the central axis. Two parabolic mirrors surround the magnetic field and reflect the incident ADM-induced radiation to the receivers at the focal point. This scheme is similar to the BRASS experiment~\cite{BRASShppt}.}
    \label{fig:scheme1}
\end{figure}

In the dish-antenna or cylindrical metal barrel experiment, the emitted power from the dish surface is $P=\frac{1}{2}\rho_a g_{a\gamma}^2  B_{\rm e}^2 \frac{A_{\rm dish}}{m_a^2}$~\cite{BREAD:2021tpx}. It is similar to our result $P=\frac{1}{2}\rho_a g_{a\gamma}^2  B_{\rm e}^2 \frac{A}{m_a^2}$ in Eq.\,(\ref{pJ}) using $\langle \rm{cos^2} \rangle=\frac{1}{2}$. When the side area $A$ of the B-field parallel to $\mathbf{B}$ is equal to $A_{\rm dish}$, the two powers are the same. It seems that the photon from axion conversion is emitted from the interface between the magnetic field and the outside. But, in our scheme, when $R\omega\ll1$, the power in proportional to the square of the volume and the mass so that the maximum power $P(\omega)$ for $5\gtrsim R\gtrsim0.5\,$m appears at $0.1\lesssim\omega\lesssim1\,\mu$eV. In this sense, our scheme is more suitable to detect the axions at $0.1\lesssim\omega\lesssim1\,\mu$eV band.

Compared to the LC circuit (e.g.,\,\cite{Ouellet:2018beu}) and cavity experiments (e.g.,\,\cite{ADMX:2021nhd}), our detection scheme is simpler but requires a much larger volume of B-field. It seems more reasonable to exploit the detection of the axion-induced radiation by using some other particular experiments (e.g.,\,stellarators) with a very strong and large B-field. The main challenge is to detect the un-concentrated radiant energy in very low temperature mode.

Though the proposed detection mass in our scheme is in $1-10^4\,$neV band, it is not constrained by resonance requirements as in cavity detections. At higher frequency, the diffraction effect of radiation can be weakened so that we can focus the radiation to a small detector according to geometric optics. Therefore, it could, in principle, be used to detect axion with mass $m_a>100\,\mu$eV. Fig.\,\ref{fig:scheme1} shows the detection scheme for infrared radiated photons. Two parabolic mirrors surround the magnetic field and reflect the incident ADM-induced radiation to the receivers at the focal point. The energy flux can be calculated by Eqs.\,(\ref{equ:numSx})-(\ref{equ:numSz}).

\section{Conclusion}
\label{sec:conclusion}
In summary, we have studied the direct radiation excited by the oscillating axion DM in a homogenous magnetic field and tried to devise a scheme to detect such axions.

We have concretely derived the analytical expression of the axion-induced radiated power for cylindrical uniform magnetic field, see Eqs.\,(\ref{pow2})-(\ref{Pl}). In the long wave limit $R\omega\ll1$, the radiation power is proportional to the square of the volume and the axion mass $m_a$. When $R\omega\gtrsim1$, the target radiations oscillate and their peak powers are proportional to the side area and inversely to $m_a^2$. The maximum power locates at $\omega\sim\frac{3\pi}{4R}$ a for fixed radius $R$. Therefore, the axion in the mass range $1-10^4$\,neV is suitable for detecting for a finite volume of the magnetic field. The radiated energy flux is also derived in Eqs.\,(\ref{equ:numSx})-(\ref{equ:S0}).

Finally, we have discussed a scheme to detect the axions in this mass range using four detectors of different bandwidths surrounding the B-field, see Fig.\,\ref{fig:scheme}. Our expected sensitivity under the typical parameter values is far beyond the existing constraints except for ADMX experiment, see Fig.\,\ref{fig:sensitivity}.

More advanced and realistic detection schemes and techniques can be further developed in the future, and our results are useful for detecting the ADM in strong man-made or natural magnetic fields.

\begin{acknowledgements}
We thank Seishi\,Enomoto, Chengfeng\,Cai, and Yi-Lei\,Tang for useful discussions and comments.
This work is supported by the National Natural Science Foundation of China (NSFC) under Grant No. 11875327, the Fundamental Research Funds for the Central Universities, China, and the Sun Yat-sen University Science Foundation.
\end{acknowledgements}

\end{document}